\documentclass[%
reprint,
amsmath,amssymb,
aps,
superscriptaddress,
prc
]{revtex4-1}

\usepackage{graphicx}
\usepackage{dcolumn}
\usepackage{bm}
\usepackage{hyperref}

\usepackage[capitalise]{cleveref}
\usepackage{comment}

\begin{document}
	
	
	\title{The universal scaling of kinetic freeze-out parameters across different collision systems at the LHC energy}
	
	\author{Lian Liu}
	\affiliation{School of Mathematics and Physics, China University of
		Geosciences (Wuhan), Wuhan 430074, China}
	\author{Zhong-Bao Yin}
	\affiliation{Key Laboratory of Quark and Lepton Physics (MOE) and Institute
		of Particle Physics, Central China Normal University, Wuhan 430079, China}
	\author{Liang Zheng}\email{zhengliang@cug.edu.cn}
	\affiliation{School of Mathematics and Physics, China University of
		Geosciences (Wuhan), Wuhan 430074, China}
	\affiliation{Key Laboratory of Quark and Lepton Physics (MOE) and Institute
		of Particle Physics, Central China Normal University, Wuhan 430079, China}

	\date{\today}
	
	\begin{abstract}
		In this paper, we perform the Tsallis Blast-Wave analysis on the transverse momentum spectra of identified hadrons produced in a wide range of collision systems at the Large Hadron Collider (LHC) including pp, pPb, XeXe and PbPb collisions. The kinetic freeze-out properties are investigated across these systems varying with the event multiplicity. We find that the extracted kinetic freeze-out temperature, radial flow velocity and the non-extensive parameter exhibit a universal scaling behavior for these systems with very different geometric size, especially when the independent baryon Tsallis non-extensive parameter is considered. This universality may indicate the existence of a unified partonic evolution stage in different collision systems at the LHC energies. 
		
	\end{abstract}
	

	\maketitle
	
	
	\section{Introduction}
	\label{sec:level1}

	It is indicated by lattice quantum chromodynamics (QCD) calculations that a deconfined quark-gluon plasma (QGP) state of nuclear matter might exist with high enough temperatures and densities~\cite{Broniowski:2008vp}. These extreme conditions are believed to be achieved in relativistic nucleus nucleus (AA) collisions, in which the QCD parton degrees of freedom are released from the nucleons and interact with each other creating the thermalized QGP medium. The deconfined parton matter expands rapidly under the thermal pressure against the surrounding vacuum until the temperature of the system drops below which the partons are converted to hadrons, developing substantial expansion flow in the parton evolution stage~\cite{Heinz:2004qz}. After further rescatterings between the produced hadronic objects for a while, the system becomes so dilute that all particle interactions are ceased~\cite{Retiere:2003kf}. The evolving information of the medium usually described by the relativistic fluid hydrodynamics~\cite{Heinz:2013th,Gale:2013da} is thus encoded in the momentum distributions of the final state particles. The transverse momentum ($p_T$) spectra of identified hadrons can be utilized to extract the system properties even in the early stage of the evolution within the Blast-Wave global analysis framework~\cite{Schnedermann:1993ws,Schnedermann:1994gc}. In the Blast-Wave model, particle spectra are obtained within a hydrodynamic framework through the thermal particle emissions from a freeze-out surface of the flowing medium with the kinetic freeze-out temperature $T$ and the radial expansion flow velocity profile $\beta(r)$. 
	
	The Boltzmann-Gibbs Blast-Wave (BGBW) model has been widely used to study the $p_T$ distributions of identified hadrons in heavy ion collisions. This approach assumes local equilibrium of the system so that a Boltzmann distribution can be applied to describe the emitting particles in the local rest frame of the expanding fluid. Considering the equilibrium assumption generally fails at high $p_T$, this BGBW framework is believed to work only in the low $p_T$ region and sensitive to the choice of the fitted $p_T$ range~\cite{STAR:2008med,ALICE:2013mez}. The Blast-Wave model is further developed with the inclusion of the Tsallis statistics~\cite{Tsallis:1999nq}, by assuming the Tsallis distributions for the emitting particles rather than the exponential type distributions used in the BGBW model. The Tsallis Blast-Wave (TBW) model is expected to account for the non-equilibrium or hard scattering effects by incorporating the non-extensive parameter $q$ and describe the final state hadron spectra for an extended $p_T$ range~\cite{De:2007zza,Wilk:2008ue,Alberico:1999nh,Osada:2008sw,Biro:2003vz,Bhattacharyya:2015hya}. Being successful in understanding a wide range of complex systems and utilized to study various high energy collisions~\cite{Urmossy:2011xk,Urmossy:2014gpa,Shao:2009mu}, the physical interpretation of the non-extensive parameter q in Tsallis statistical method is still under discussion. It is found that the hard scattering process in high energy collisions can deliver a power law distribution of high $p_T$ hadrons from jet fragmentations with the power index related to the non-extensive parameter $q$~\cite{Wong:2015mba}. This $q$ parameter characterizes the degree of deviation from the equilibrium assumption. If $q$ approaches unity, the TBW model returns to the BGBW model.

	The energy dependence of the radial flow and the kinetic freeze-out temperature has been systematically studied~\cite{Shao:2009mu,Tang:2011xq, Chen:2020zuw,Lao:2016gxv,Lao:2017skd,Zhang:2014jug}
	With the unique capabilities to account for the non-equilibrium effects, the TBW model is also believed to be useful in describing the hadron productions in small systems like proton proton (pp), proton nucleus (pA) and peripheral AA collisions~\cite{Jiang:2013gxa,Tang:2008ud,Che:2020fbz}. 
	The recent collectivity like behaviors observed in small collision systems receives a lot of discussions on whether the QGP matter is formed in these collisions~\cite{Nagle:2018nvi,Adolfsson:2020dhm,CMS:2010ifv,CMS:2016fnw,ALICE:2016fzo,ALICE:2015ial,ALICE:2018pal}. The universal strangeness enhancement effect scaled with the event multiplicity found at the LHC energy suggests a unified mechanism might induce the enhanced multi-strange hadron productions in both small and large systems~\cite{ALICE:2017jyt,Acharya:2019kyh,Adam:2015vsf}. Searching for the universality in other collectivity related effects like the radial expansion flow velocity and the freeze-out temperature across different collision systems and studying its system size dependence would be important to understand the appearance of the QGP like effects in small systems. 
	It is speculated that initial energy density induced hot spots caused by Color Glass Condensate formalism may generate strong collective flow and sizable temperature fluctuations especially in small systems. The imprints of the initial fluctuations survived in the final state hadron momentum spectra at low and intermediate $p_T$ region are expected to be captured in the TBW model with the non-extensive parameter $q$. The temperature and flow velocity extracted from the TBW model are correlated with the $q$ parameter by the shear and bulk viscosity in linear and quadratic forms~\cite{Wilk:1999dr,Wilk:2008ue}. Investigating the kinetic freeze-out features dependent on the event multiplicity may shed a light on the understanding of the origin of the collectivity like behavior observed in small systems.
	
	In this work, we will use the Tsallis Blast-Wave model to fit the transverse momentum spectra of identified hadrons $\pi^{\pm},K^{\pm},p,\bar{p}$ produced in $\sqrt{s}=7$ and $13$~TeV pp collisions~\cite{ALICE:2018pal,ALICE:2020nkc}, $\sqrt{s_{NN}}=5.02$~TeV pPb collisions~\cite{ALICE:2016dei}, $\sqrt{s_{NN}}=5.44$~TeV XeXe collisions~\cite{ALICE:2021lsv}, and in $\sqrt{s_{NN}}=2.76$~TeV and $5.02$~TeV PbPb collisions~\cite{ALICE:2015dtd,ALICE:2019hno}. The extracted parameters $T$, $\beta$ and $q$ will be systematically compared across different systems with and without considering the separate non-equilibrium parameter $q$ for baryons.
	The rest of this paper is organized as follows: we illustrate the implementation of the Tsallis Blast-Wave model approach in Sec.~\ref{sec:formalism}. The results of the extracted kinetic freeze-out parameters are presented in Sec.~\ref{sec:results}. The major conclusions and discussions are summarized in Sec.~\ref{sec:summary}.

	\section{\label{sec:formalism}The Tsallis Blast-Wave model}
	The BGBW model derived from the hydrodynamic framework has been widely used to fit the transverse momentum distribution data in heavy ion collisions. It is well known that the BGBW model follows the local thermal equilibrium assumption and only works in the low $p_T$ region of the momentum spectra. The Tsallis distribution smoothly connects the power law type jet induced high $p_T$ part and the exponential type hydrodynamics dominant low $p_T$ region of the momentum distributions. With the inclusion of the non-extensive parameter, the TBW model is expected to describe the evolution from pp collisions to the central AA collisions in a consistent way. It is straightforward to implement the TBW model by replacing the particle emission sources in the local rest frame of the fluid cell from the Boltzmann distribution to the Tsallis distribution.  
	
	The invariant differential particle yield for a hadron with mass $m$ in TBW model can be written in the form of 
	\begin{eqnarray}\label{eq:TBW}
	\frac{d^{2}N}{2\pi m_{T}dm_{T}dy}|_{y=0} & = & A\int^{+y_{b}}_{-y_{b}}m_{T}\cosh(y_{s})dy_{s}\int^{\pi}_{-\pi}d\phi \\ \nonumber
	& \times & \int^{R}_{0}rdr[1+\frac{q-1}{T}(m_T\cosh(y_s)\cosh(\rho) \\ \nonumber
	& &-p_T\sinh(\rho)\cos(\phi))]^{-1/(q-1)}.
	\end{eqnarray}
	$T$ is the global temperature of the expanding thermal sources from which particles are emitted. $R$ is the hard sphere edge along the transverse radial direction.
	$A$ denotes the normalization constant.
	$m_{T}=\sqrt{p^{2}_{T}+m^{2}}$ is the transverse mass of a particle.
	$y_{s}$ represents the source rapidity. $y_{b}$ gives the beam rapidity.
	$\phi$ is the particle emission angle with respect to the flow velocity.
	 $\rho=\tanh^{-1}\beta(r)$ is the radial flow profile obtained in a self-similar way with the transverse flow velocity parametrized as $\beta(r)=\beta_{S}(\frac{r}{R})^n$ under the longitudinal boost invariant assumption. $\beta(r)$ is defined in the range $0\leqslant r\leqslant R$ with the surface velocity $\beta_{S}$ at the edge of the fireball hard sphere $R$ and the flow profile index $n$.
	The average transverse flow velocity can be expressed as $\langle\beta\rangle=\beta_{S}\cdot2/(2+n)$. The relevant model parameters can be extracted based on Eq.~\ref{eq:TBW} using the least $\chi^{2}$ method.
	
	In this study, the value of $n$ is fixed to 1 for the consideration of a linear velocity profile, thus one gets $\langle\beta\rangle=2/3\beta_{S}$.
	It has been argued that in small collision systems, baryon number might play an important role in hadron production and the characteristic grouping of meson and baryon sector are needed to provide better description to the $p_T$ spectra in TBW fits~\cite{Tang:2008ud,Jiang:2013gxa}. In the rest of this paper, we refer to the fits including all of the mesons and baryons with a combined non-extensive parameter $q$ as the default TBW fit and those with four fit parameters using different $q$ for meson ($q_M$) and baryon ($q_B$) as the TBW4 fit.
	In this paper, we perform both the TBW and TBW4 fits to the charged pion, kaon, and proton $p_T$ spectra at the LHC energies measured by the ALICE collaboration to extract the kinetic freeze-out parameters. 
	
	\section{\label{sec:results}Results}
	\subsection{Transverse momentum spectra}
	
	This section compares TBW and TBW4 model fits of the transverse momentum spectra for charged pion, kaon and proton in different collision systems at the LHC energies. We restrict ourselves to the comparable spectra range $p_T<3$ GeV$/c$ across all the collision systems to extract the bulk features. The average flow velocity $\langle\beta\rangle$ is required to be less than $2/3$ and greater than $0$ during the fit procedures to achieve a better convergence and get rid of the non-physical parameter space. For some of the very peripheral collisions, the flow parameters are very close to the boundary. We fix the flow velocity parameter $\langle\beta\rangle=0$ to reduce the uncertainty of the fits for those peripheral bins. It has been verified that fixing the flow velocity to zero in these peripheral bins does not change the values of the other extracted model parameters. 
	
	\begin{figure*}[htbp]
		\begin{center}
			\includegraphics[width=1.0\textwidth]{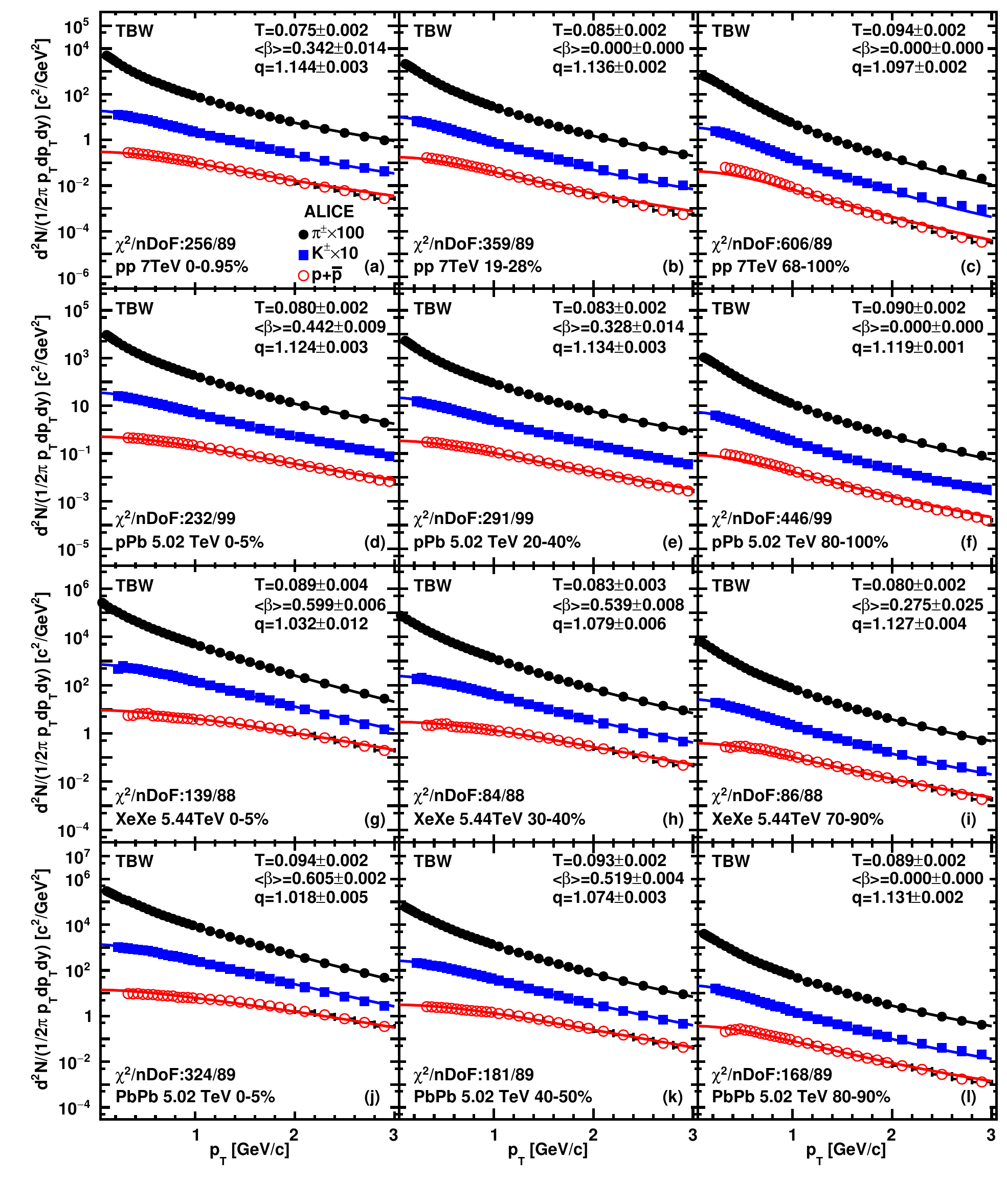}
			\caption{The default TBW fits to hadron spectra in pp collisions at $\sqrt{s}=$ 7~TeV, pPb collisions at $\sqrt{s_{\rm{NN}}}=$ 5.02~TeV, XeXe collisions at $\sqrt{s_{\rm{NN}}}=$ 5.44~TeV and PbPb collisions at $\sqrt{s_{\rm{NN}}}=$ 5.02~TeV from top to bottom panels. Results from the central, semi-central and peripheral collisions are shown in the left column, middle column and right column, respectively. The markers represent  ALICE experimental data\cite{ALICE:2018pal,ALICE:2020nkc,ALICE:2016dei,ALICE:2021lsv,ALICE:2015dtd,ALICE:2019hno} of identified particle species. Uncertainties on experimental data represent quadratic sums of statistical and systematic uncertainties. The solid curves represent fit results from the TBW model.}\label{fig:spectra_TBW}
		\end{center}
	\end{figure*}

	\begin{figure*}[htbp]
		\begin{center}
			\includegraphics[width=1.0\textwidth]{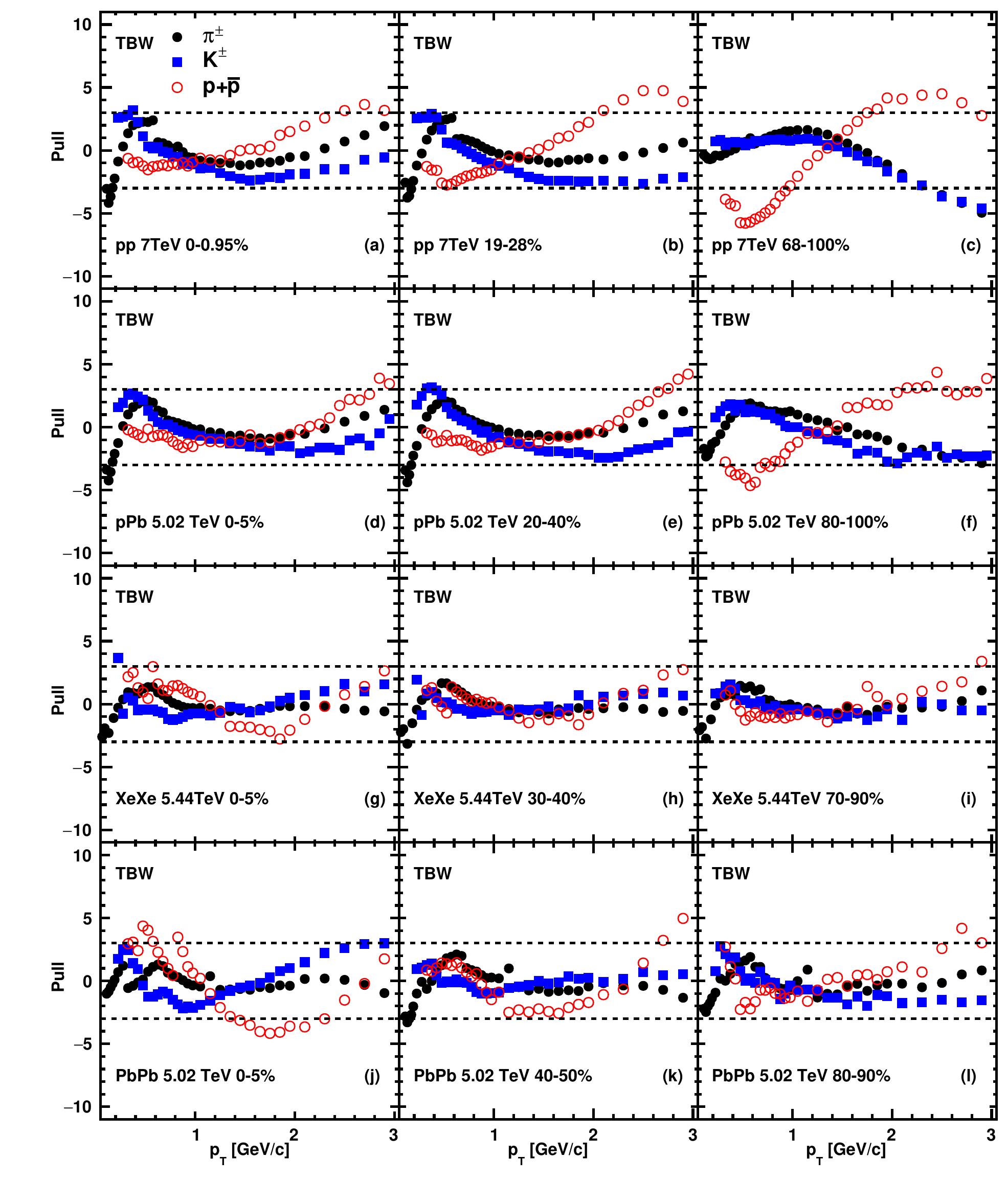}
			\caption{The deviations of TBW model fits to hadron spectra divided by data uncertainties  in pp collisions at $\sqrt{s}=$ 7~TeV, pPb collisions at $\sqrt{s_{\rm{NN}}}=$ 5.02~TeV, XeXe collisions at $\sqrt{s_{\rm{NN}}}=$ 5.44~TeV and PbPb collisions at $\sqrt{s_{\rm{NN}}}=$ 5.02~TeV from top to bottom panels. The markers represent the results for different particle species. The dashed lines represent where the difference between model and experiment data is three times the error of data.}\label{fig:pull_TBW}
		\end{center}
	\end{figure*}
	
	\begin{figure*}[htbp]
		\begin{center}
			\includegraphics[width=1.0\textwidth]{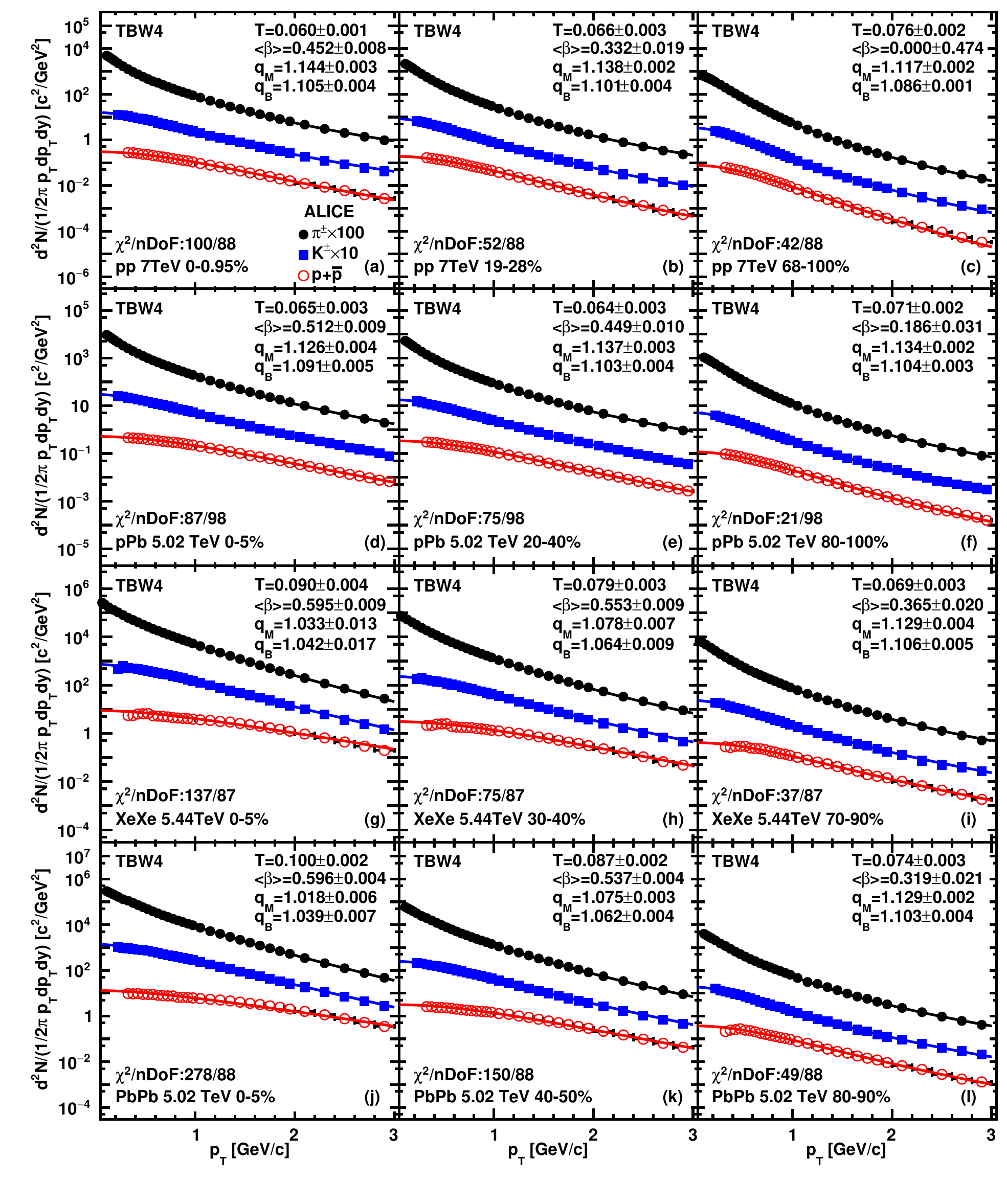}
			\caption{The default TBW4 fits to hadron spectra in pp collisions at $\sqrt{s}=$ 7~TeV, pPb collisions at $\sqrt{s_{\rm{NN}}}=$ 5.02~TeV, XeXe collisions at $\sqrt{s_{\rm{NN}}}=$ 5.44~TeV and PbPb collisions at $\sqrt{s_{\rm{NN}}}=$ 5.02~TeV from top to bottom panels. Results from the central, semi-central and peripheral collisions are shown in the left column, middle column and right column, respectively. The markers represent ALICE experimental data\cite{ALICE:2018pal,ALICE:2020nkc,ALICE:2016dei,ALICE:2021lsv,ALICE:2015dtd,ALICE:2019hno}  of identified particle species. Uncertainties on experimental data represent quadratic sums of statistical and systematic uncertainties. The solid curves represent fit results from the TBW4 model.}\label{fig:spectra_TBW4}
		\end{center}
	\end{figure*}
	
	\begin{figure*}[htbp]
		\begin{center}
			\includegraphics[width=1.0\textwidth]{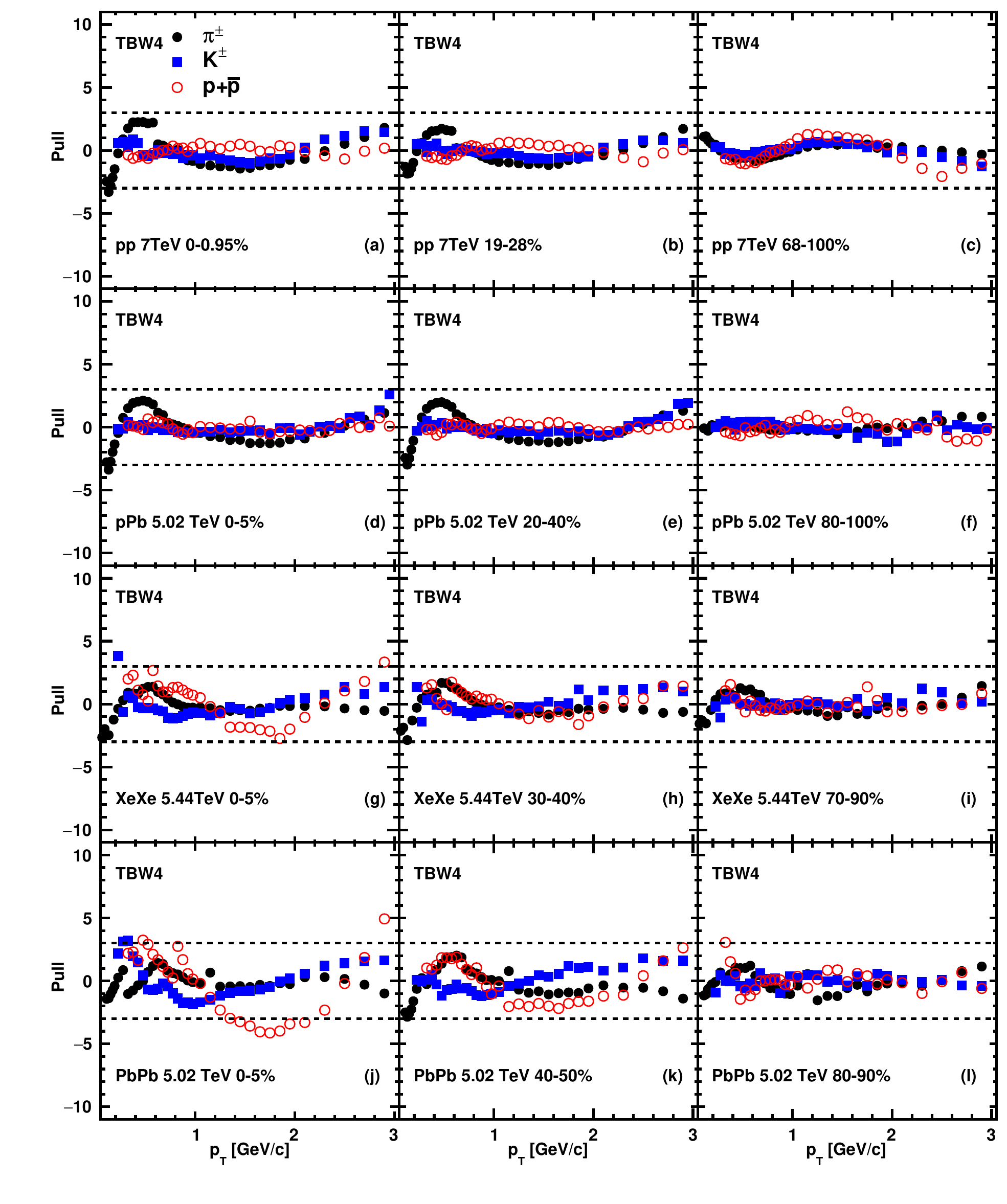}
			\caption{The deviations of TBW4 model fits to hadron spectra divided by data uncertainties  in pp collisions at $\sqrt{s}=$ 7~TeV, pPb collisions at $\sqrt{s_{\rm{NN}}}=$ 5.02~TeV, XeXe collisions at $\sqrt{s_{\rm{NN}}}=$ 5.44~TeV and PbPb collisions at $\sqrt{s_{\rm{NN}}}=$ 5.02~TeV from top to bottom panels. The markers represent the results for different particle species. The dashed lines represent where the difference between model and experiment data is three times the error of data. }\label{fig:pull_TBW4}
		\end{center}
	\end{figure*}
	
	\begin{figure*}[htbp!]
		\begin{center}
			\includegraphics[width=1.0\textwidth]{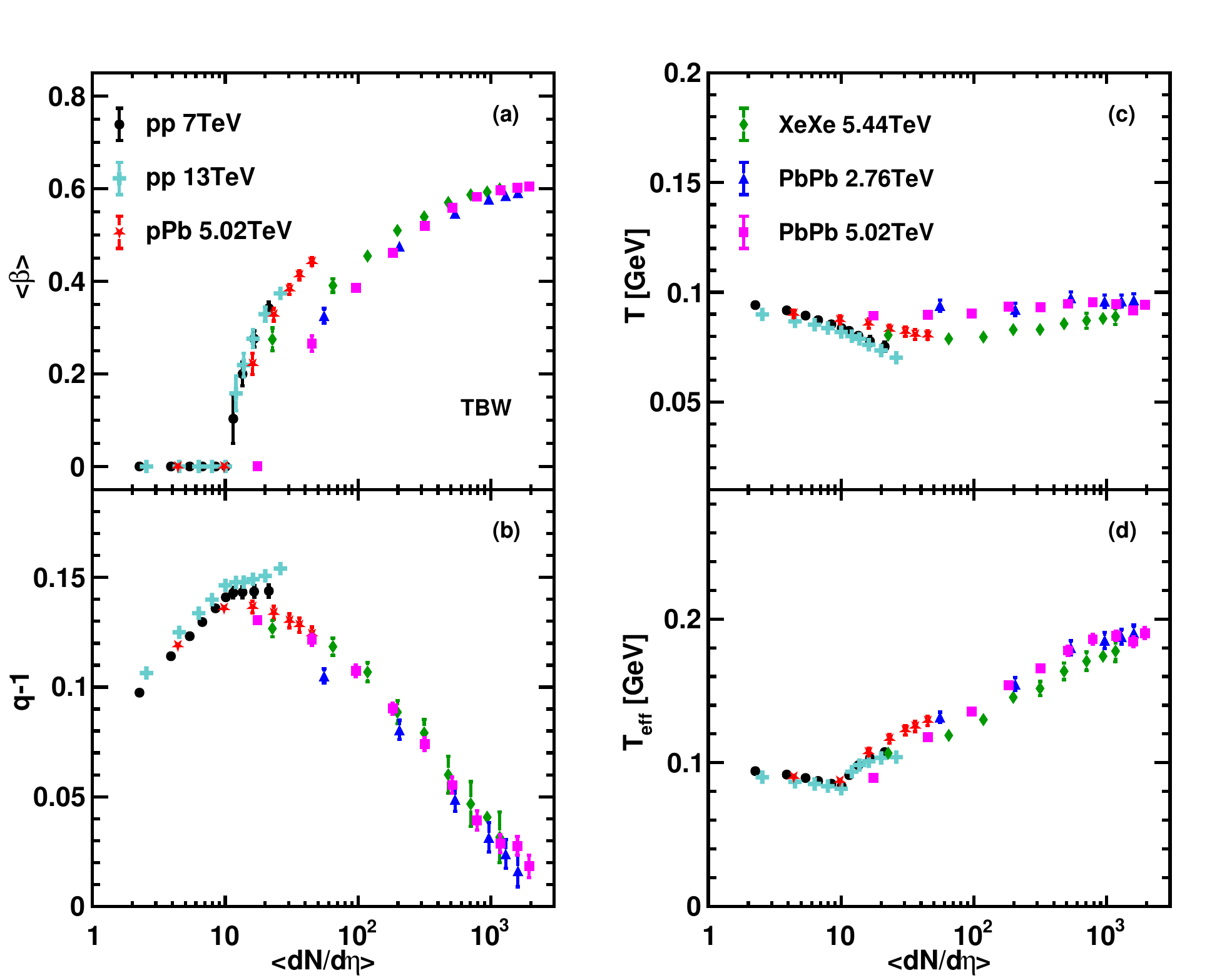}
			\caption{Charge multiplicity dependence of the extracted freeze-out parameters and the effective temperature $T_{eff}$ of different collision systems from TBW fits. Solid symbols with the same style represent different centrality classes in each collision system. 
			} \label{fig:parameterTBW}
		\end{center}
	\end{figure*}
	
	\begin{figure*}[htbp!]
		\begin{center}
			\includegraphics[width=1.0\textwidth]{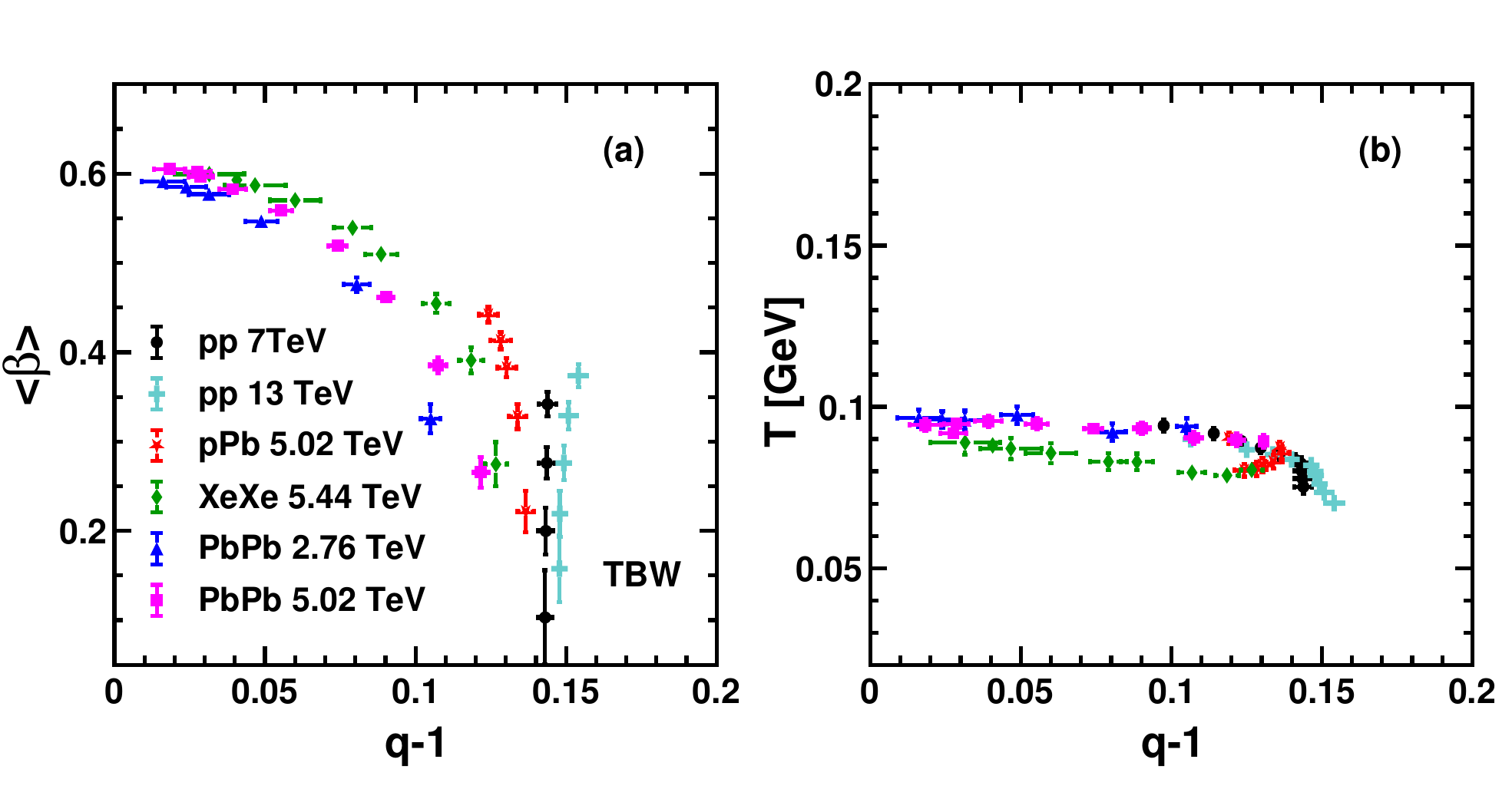}
			\caption{$\langle\beta\rangle$ vs $q-1$ and $T$ vs $q-1$ of different collision systems from TBW fits. Solid symbols with the same style represent different centrality classes in each collision system. 
			} \label{fig:parameter_Cov_TBW}
		\end{center}
	\end{figure*}

	\begin{figure*}[htbp!]
		\begin{center}
			\includegraphics[width=1.0\textwidth]{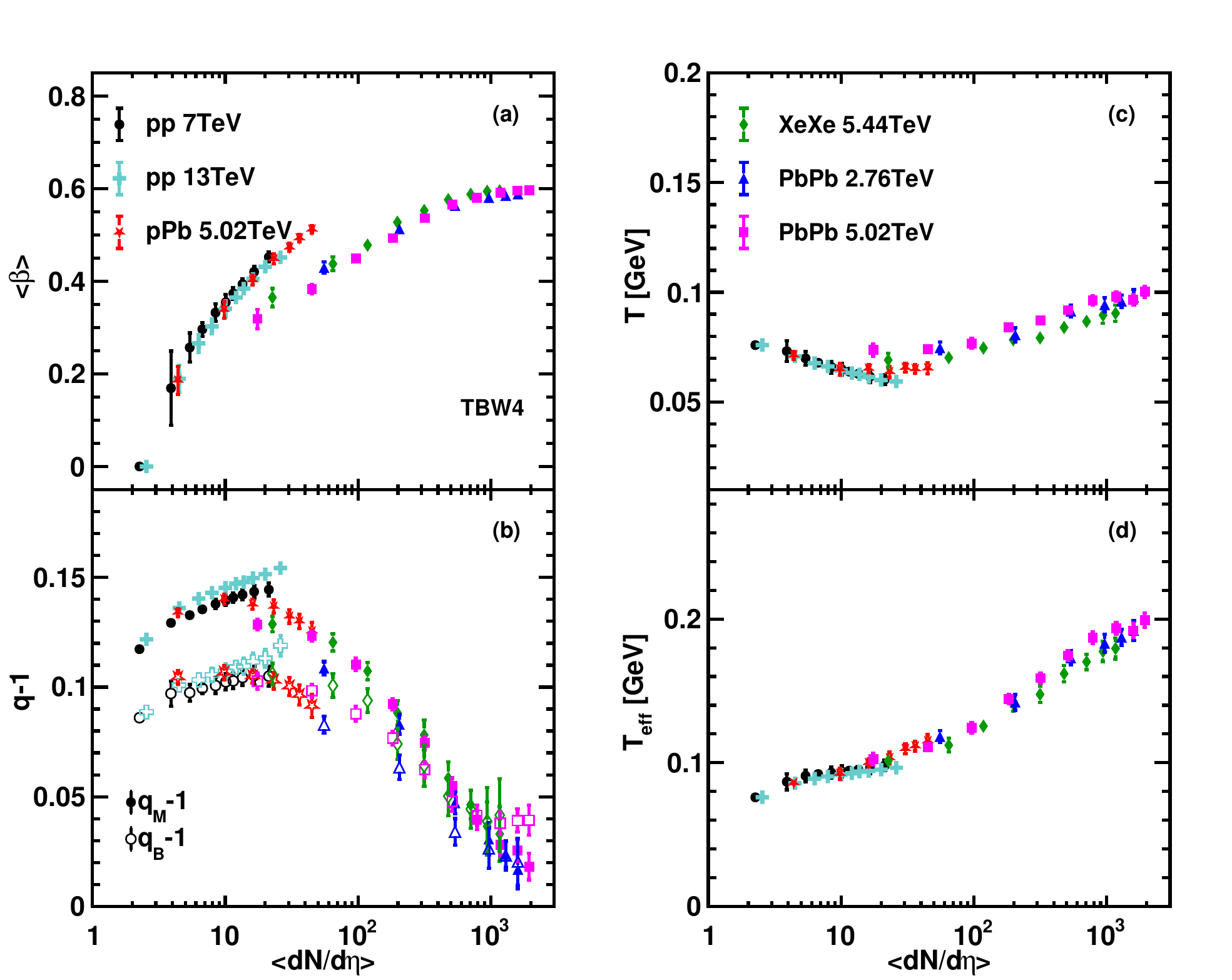}
			\caption{Charge multiplicity dependence of the extracted freeze-out parameters and the effective temperature $T_{eff}$ of different collision systems from TBW4 fits. Solid symbols with the same style represent different centrality classes in each collision system. In panel (c), open markers represent the results of $q_B$ and solid markers represent the results of $q_M$. 
			} \label{fig:parameterTBW4}
		\end{center}
	\end{figure*}
	
	\begin{figure*}[htbp!]
		\begin{center}
			\includegraphics[width=1.0\textwidth]{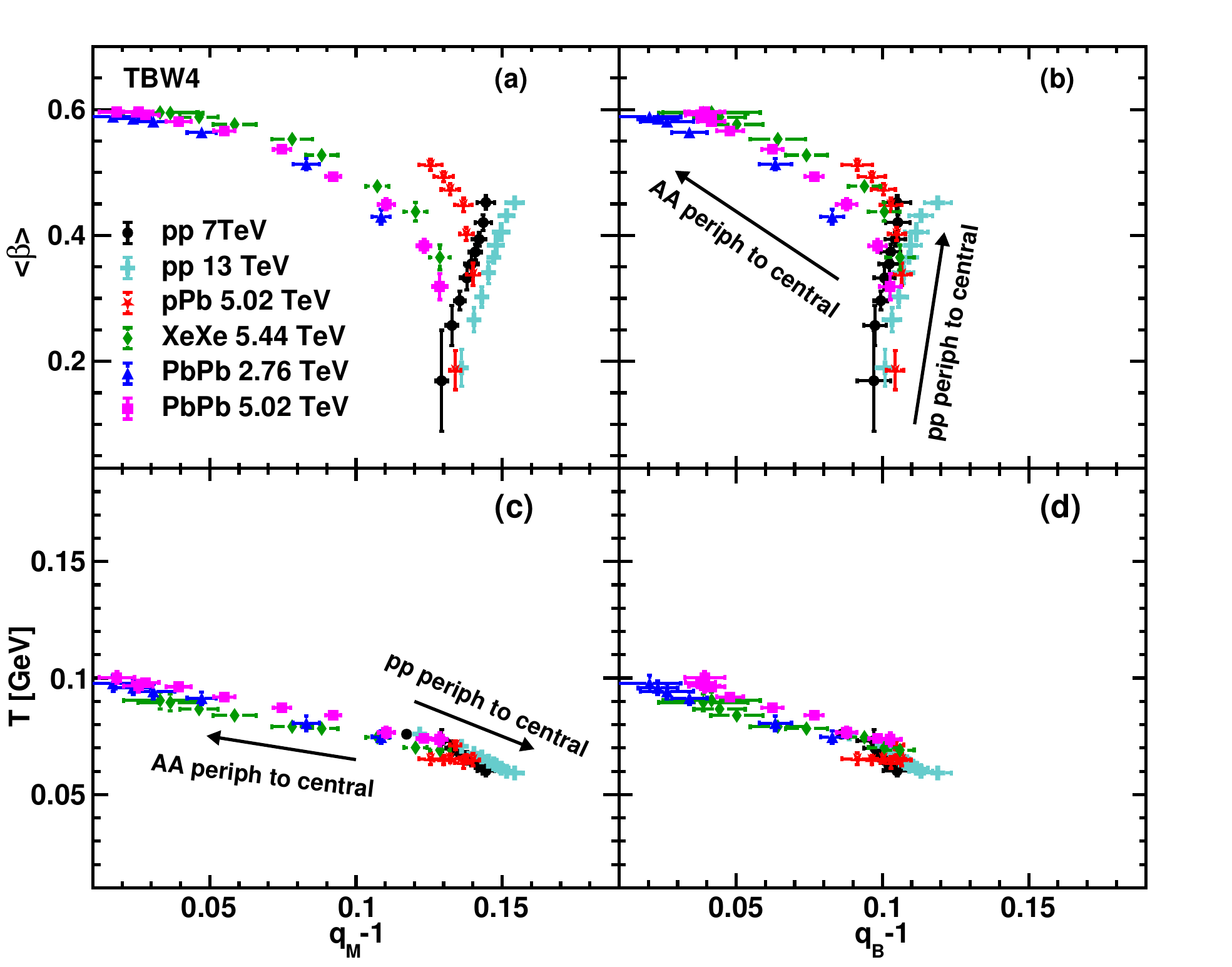}
			\caption{$\langle\beta\rangle$ vs $q-1$ and $T$ vs $q-1$ of different collision systems from TBW4 fits for meson non-extensive parameter (left column) and baryon non-extensive parameter (right column). Solid symbols with the same style represent different centrality classes in each collision system. Arrows indicate the direction from peripheral to central classes for AA and pp collisions in the parameter space.
			} \label{fig:parameter_Cov_TBW4}
		\end{center}
	\end{figure*}

	We present the transverse momentum spectra fits based on the default TBW model analysis in Fig.~\ref{fig:spectra_TBW}. Only examples for four collision systems from central to peripheral centralities are demonstrated in this figure, more details about the extracted fitting parameter values can be found in Tab.~\ref{tabTBW}. The black solid circles, blue solid squares and red open circles represent the experimental data in each panel. The solid lines mark the corresponding fit functions to each particle species. The results of the extracted model parameters and the $\chi^2/nDoF$ values are also displayed in the figure. It is shown that a stronger radial flow is expected in central collisions compared to that in peripheral collisions for all systems. The freeze-out temperature mildly changes with the centrality in each collision system. The non-extensive parameter grows from peripheral to central collisions in pp collisions but drops very fast in AA collisions approaching 1 when the system is reaching equilibrium. The $\chi^2/nDoF$ values become quite large in both very small systems like pp collisions at $\sqrt{s}=7$ TeV and very large systems like PbPb collisions at $\sqrt{s_{NN}}=5.02$ TeV. The deviations of the fits to experimental data divided by experimental uncertainties (usually defined as $pull=(fit-data)/(data\ error)$) are shown in Fig.~\ref{fig:pull_TBW}. The pull distributions can be used to quantify the agreement between the model fits and the experimental data. In all systems, the fit results seem to be consistent with data in the intermediate $p_T$ region around 1.5 GeV$/c$, but sizable deviations can be found generally in both smaller and higher $p_T$ regime. It is speculated that the deviations at low $p_T$ part may come from contributions due to the resonance decay effects. 
	
	For a comparison, we also show the transverse momentum spectra and pull distributions based on the TBW4 fits in Fig.~\ref{fig:spectra_TBW4} and Fig.~\ref{fig:pull_TBW4}, more details about the extracted fitting parameters can be found in Tab.~\ref{tabTBW4}. Modifying the TBW model to enable the independent $q_B$ parameter for baryons has reached high quality fits for small systems, while the improvement to large collision systems are quite limited. The TBW4 fits are found to work significantly better especially in pp and pPb collisions while $\chi^2/nDoF$ values in central XeXe and PbPb collisions shown in Fig.~\ref{fig:spectra_TBW4} are quite similar to those obtained in TBW fits presented in Fig.~\ref{fig:spectra_TBW}. The improvement in small systems mainly comes from a better description to the proton $p_T$ distributions as is shown in Fig.~\ref{fig:pull_TBW4}. This is expected due to the inclusion of the baryon grouped non-extensive parameter $q_B$. This improvement indicates the importance of baryon number in the fragmentation process of small systems. However, it is found that the pull distributions are not flattened to zero especially for protons in large systems even with the independent $q_B$ considered. There is a possibility that in central AA collisions, nuclear medium modifications like shadowing effects and parton energy loss effects can be important to understand this discrepancy. From Fig.~\ref{fig:spectra_TBW4}, it is also found that the extracted freeze-out temperature is smaller than that in the default TBW fits for pp and pPb collisions. The non-extensive parameter for meson is usually larger than that for baryon except in the central AA collisions when the off equilibrium effects are vanishing. These two non-extensive parameters $q_M$ and $q_B$ become similar in central XeXe and PbPb collisions.

	\subsection{\label{sec:par}Kinetic freeze-out parameters}
	
	The model parameters extracted from the TBW and TBW4 fits dependent on the event multiplicity are studied in this section. We present the results of kinetic freeze-out temperature $T$, average radial flow velocity in the transverse plane $\langle\beta\rangle$ and non-extensive parameter $q$ extracted from the TBW analysis varying with the mid-rapidity charged particle multiplicity in Fig.~\ref{fig:parameterTBW}. The charged multiplicity is quantified with the average charged particle number within $|\eta|<0.5$ per unit pseudorapidity $\langle dN/d\eta \rangle$ in each event class. The black circles, cyan crosses, red stars, green diamonds, blue triangles and magenta squares represent the results of pp 7TeV, pp 13TeV, pPb 5.02 TeV, XeXe 5.44 TeV, PbPb 2.76 TeV and PbPb 5.02 TeV, respectively.
	It is shown in Fig.~\ref{fig:parameterTBW}(a) that the multiplicity dependent flow velocity can be divided into two groups, one group contains mainly large symmetric systems consisting of XeXe 5.44 TeV, PbPb 2.76 TeV and PbPb 5.02 TeV, the other group includes the small systems ranging from pp 7 TeV, pp 13 TeV to pPb 5.02 TeV. The average flow velocity grows with the multiplicity rapidly in the low $\langle dN/d\eta\rangle$ region but saturates at about 0.6 when the average charge number density $\langle dN/d\eta \rangle $ approaches 1000. The onset of non-zero radial flow starts from $\langle dN/d\eta \rangle $ around 10 to 20 in two groups. The branching of small system flow velocity with respect to the large system flow at the same $\langle dN/d\eta\rangle$ can be understood in the context of parton density difference in the transverse plane. Higher initial densities in pp and pA collisions might produce larger pressure against the surrounding environment~\cite{Shuryak:2013ke,Castorina:2019vex}.
	
	The non-extensive parameter is shown in Fig.~\ref{fig:parameterTBW}(b) as $(q-1)$ versus the average charge particle density. It is found that the non-extensive parameter initially increases with the charge particle density and then drops when $\langle dN/d\eta \rangle $ becomes larger than 10. The non-extensive parameters from different collision systems seem to scale together. Deviations arise in the region of $\langle dN/d\eta \rangle $ between 10 and 50. The non-extensive parameter in pPb collisions in the high multiplicity regime follows the AA collision trend, unlike the radial flow velocity which follows the trend of pp collisions. On the other hand, the extracted kinetic freeze-out temperature $T$ shows the opposite dependence on the event multiplicity compared to the non-extensive parameter $(q-1)$. As is displayed in Fig.~\ref{fig:parameterTBW}(c), a slight decreasing within $\langle dN/d\eta \rangle <10$ and a mild growth of the kinetic freeze-out temperatures dependent on the charge particle density in the region of $\langle dN/d\eta \rangle >40$ is observed. In both non-extensive parameter and freeze-out temperature comparison, the high multiplicity pp 13 TeV results are found to slightly deviate from the global trend. It is also interesting to see that the radial flow velocity in high multiplicity pPb collisions is similar to pp collisions, while the non-extensive parameter in the same events is following the AA trend.
	
	In the end, we also present the effective temperature of the system defined as $T_{eff}=\sqrt{\frac{1+\langle\beta\rangle}{1-\langle\beta\rangle}}T$~\cite{Heinz:2004qz,Che:2020fbz} in Fig.~\ref{fig:parameterTBW}(d). The effective temperature incorporating the radial flow velocity seems to be insensitive to the collision system but only relies on the charge multiplicity. A turning point in the $T_{eff}$ distribution is seen at $\langle dN/d\eta \rangle  \sim 10$ similar to the non-extensive parameter distribution, above which the temperature increases monotonously up to $\langle dN/d\eta \rangle \sim 10^3$ in a unified way across all different systems ranging from pp collisions to AA collisions.   
	
	In the framework of non-equilibrium statistics, the temperature and the flow velocity can be related to viscosity with the linear or quadratic dependence on the non-extensive parameter $(q-1)$. The $\langle\beta\rangle$ vs $(q-1)$ and $T$ vs $(q-1)$ distributions from TBW fits are shown in Fig.~\ref{fig:parameter_Cov_TBW}. The non-zero radial flow velocity and $(q-1)$ can be roughly described by a universal quadratic dependence, with the exception that high multiplicity pp collisions are more likely to have larger non-extensive parameter instead of having vanishing $(q-1)$ value as seen in central AA collisions. The kinetic freeze-out temperature $T$ vs $(q-1)$ also shows little sensitivity to the collision system, while a slightly diverged branch consisting of XeXe collisions and pPb collisions is observed. The deviation of the XeXe and pPb results compared to the large symmetric systems is possibly coming from the fluctuation effects in the initial conditions whose initial spatial geometries can vary significantly due to the non-spherical shape of Xe nuclei and the asymmetric colliding nuclei involved in the pPb collisions. The potentially produced hot spots region during the fluctuations may lead longer life time of the evolving medium and a lower freeze out temperature. 
	
	The extracted freeze-out parameters are also studied in the case with two independent non-extensive parameters $q_M$ for meson and $q_B$ for baryon. Compared to the results shown in Fig.~\ref{fig:parameterTBW}, much better scaling features are observed in Fig.~\ref{fig:parameterTBW4} for TBW4 fits. The radial flow velocity distribution is still divided into two groups with a quite early onset of non-zero radial flow in $\langle dN/d\eta\rangle$ as shown by Fig.~\ref{fig:parameterTBW4}(a). The non-extensive parameters $q_B$ for the baryon are represented by the hollow markers and compared to $q_M$ in Fig.~\ref{fig:parameterTBW4}(b). $q_M$ is found to be larger than $q_B$ especially in the low multiplicity regime. The separated meson and baryon non-extensive parameters seem to automatically converge in the high multiplicity region. This indicates that the baryon number is mostly important for the hadronization process in the small systems. For large systems, the non-equilibrium effects become negligible and the non-extensive parameter turns out to be independent of the hadron species. The radial flow velocity and the non-extensive parameter in high multiplicity pPb collisions show the same diverged features similar to the finding in Fig.~\ref{fig:parameterTBW}, following the pp collision behavior in $\langle\beta\rangle$ and the AA collision behavior in $(q-1)$. This deviation arises since the transverse overlap area in small systems is much smaller than that in large systems at the same event multiplicities~\cite{Loizides:2017ack}. The initial entropy density, represented by the particle density per unit transverse overlap area, in small systems is thus significantly enhanced, which leads to the stronger collective expansion in high multiplicity pPb collisions in contrast to that in peripheral PbPb collisions. The extracted radial flow velocity is expected to scale with the charged particle density per unit transverse overlap area across the large and small systems in a more coherent way~\cite{Petrovici:2018mpq}.
	
	A two staged temperature dependence varying with $ \langle dN/d\eta \rangle$ is found in Fig.~\ref{fig:parameterTBW4}(c) and Fig.~\ref{fig:parameterTBW4}(d) similar to the results shown in Fig.~\ref{fig:parameterTBW}(c) and Fig.~\ref{fig:parameterTBW}(d). The deviation of the temperature from PbPb collisions to XeXe collisions is less important in TBW4 fits than that in the TBW case. Without the initial decreasing part, $T_{eff}$ in TBW4 grows all the way from low to high multiplicity only with different slopes in the two regions separated by $\langle dN/d\eta \rangle \sim20$. 
	
	It is noteworthy that a turning behavior exists in the multiplicity dependence of the freeze-out parameters $\langle\beta\rangle$, $q$ and $T$ around $\langle dN/d\eta\rangle \sim $ 10 to 15 shown both in Fig.~\ref{fig:parameterTBW} and Fig.~\ref{fig:parameterTBW4}. Considering that the hadronic interactions are believed to be dominant in the low multiplicity events and the parton interactions become more important in the high multiplicity limit, the region in between may suffer from the mixture of the two types of contributions. The emergence of this feature suggests that the dominant physics mechanism dictating the evolution of the system changes from the hadron gas rescatterings to the deconfined quark gluon matter interactions. This turning behavior can be regarded as a signature that the quark and gluon degrees of freedom begin to take over in the high multiplicity region.

	Considering the TBW4 fits generally give better description to the experimental data, we also examine the $\langle\beta\rangle$ vs $(q-1)$ and $T$ vs $(q-1)$ in Fig.~\ref{fig:parameter_Cov_TBW4} from TBW4 analysis. The large system $\langle\beta\rangle$ vs $(q-1)$ relationship can be roughly described by a quadratic fit, while more complicated structures can be found for the smaller systems as displayed in Fig.~\ref{fig:parameter_Cov_TBW4}(a) and Fig.~\ref{fig:parameter_Cov_TBW4}(b). Similar to the findings in Fig.~\ref{fig:parameterTBW4}(b), central pp collisions have large $q$ and sizable flow, while large collision systems tend to have smaller $q$ and large $\langle\beta\rangle$ in central events. The parameters of peripheral pPb events are close to the pp case but become similar to the 
	AA case in high multiplicity pPb collisions. $T$ vs $(q-1)$ relationships shown in Fig.~\ref{fig:parameter_Cov_TBW4}(c) and Fig.~\ref{fig:parameter_Cov_TBW4}(d) are found to be universal for all collision systems. Central pp collisions approach large $q$ and small $T$ while central AA collisions are found to have large $T$ and small $q$. It is interesting to see that both the AA system and the pp system have similar model parameter values in the peripheral collisions. These two types of collision systems are following two distinguished evolution curve on the $\langle\beta\rangle$ vs $(q-1)$ plane and moving towards opposite direction along a universal curve in the $T$ vs $(q-1)$ space. The difference between the large systems and the small systems in the $\langle\beta\rangle$ vs $(q-1)$ space can be partly understood after considering the striking $\langle\beta\rangle$ gap between the two types of systems shown in Fig.~\ref{fig:parameterTBW}(a) and Fig.~\ref{fig:parameterTBW4}(a). Additionally, the initial spatial fluctuation effects can be also important to account for this behavior. The initial geometries of large systems are supposed to be controlled by the average shape of the overlap region while the sub-nucleon level fluctuation effects arise in the determination of initial densities for the high multiplicity small systems~\cite{Schenke:2014zha,Zheng:2021jrr}. The sizable non-equilibrium effects induced by the strong fluctuations in central pp collisions lead to increasing non-extensive parameter, unlike the central AA collisions in which $q$ is vanishing as the system approaches equilibrium. As the dynamical fluctuations in freeze-out temperature induced by the system size effects are encoded in the non-extensive parameter~\cite{Wilk:2008ue,Bozek:2012fw,Basu:2016ibk}, the variations in temperature are compensated by the corresponding non-extensive parameters, universal scalings are observed in the $T$ vs $(q-1)$ space. It is expected that the pp collisions and pPb collisions with extremely high multiplicities will approach the environment created in AA collisions, indicated by the behavior of $\langle\beta\rangle$ and $(q-1)$ in Fig.~\ref{fig:parameterTBW} and Fig.~\ref{fig:parameterTBW4}. Examining whether the turning behavior shown in Fig.~\ref{fig:parameter_Cov_TBW4}(a) and (b) exists in the ultra-central pp and pPb collisions is an interesting analysis with the future experiment data.

	\section{Summary}
	\label{sec:summary}
	In this paper, we have used the Tsallis Blast-Wave model with and without independent baryon non-extensive parameters to fit the transverse momentum spectra of charged pion, kaon and protons produced at mid-rapidity in pp collisions at $\sqrt{s}=$ 7 and 13 TeV, pPb collisions at $\sqrt{s_{\rm{NN}}}=$ 5.02 TeV, XeXe collisions at $\sqrt{s_{\rm{NN}}}=$ 5.44 TeV, and in PbPb collisions at $\sqrt{s_{\rm{NN}}}=$ 2.76 TeV and 5.02 TeV at the LHC to extract kinetic freeze-out parameters. It is found that the introduction of $q_B$ for the baryon in the TBW4 fits improves the description of identified hadron spectra especially in pp, pA and peripheral AA collisions. The multiplicity dependence of the freeze-out properties are examined across the the largely varied collision systems. A general universal scaling of the freeze-out parameters can be observed especially in the Tsallis analysis with separate baryon non-extensive parameter. 
	
	The radial flow velocities obtained from both the TBW fits and the TBW4 fits are divided into two categories consisting of small systems and large symmetric systems, respectively. The temperatures from different collision systems are scaling with the charge multiplicity in a unified way. Similar universality is also found in the non-extensive parameter distributions with some minor deviations in the most central pp collisions. By investigating the correlation of radial flow velocity $\langle\beta\rangle$ vs $(q-1)$ and kinetic freeze-out temperature $T$ vs $(q-1)$ in the TBW4 fits with independent $q_B$ assumptions, the peripheral AA collisions and pp collisions are found to be quite similar while the two types of systems are moving towards completely different direction going to central collisions in the parameter space. The asymmetric pPb collisions are close to the pp collisions in the low multiplicity region but become AA like in the high multiplicity region. A transitional behavior can be found in the evolution of the pPb collision system.
	
	The universality of the freeze-out properties revealed in this work suggests that the Tsallis Blast-Wave analysis can be applicable to various collision systems covering a wide range of event multiplicities. This feature also indicates that the evolution properties of the small size collision systems with high multiplicities and the large systems at the LHC energy can be driven by a unified physics mechanism. A parton evolution stage with sizable collective motion might exist even in the small collision systems. Extending the study to the pp collision events with even higher charge particle density~\cite{Arkani-Hamed:2015vfh} or to the intermediate size collision systems like Oxygen-Oxygen collisions~\cite{Brewer:2021kiv} in the future experiments can be important to further understand the universality of freeze-out properties observed in different collision systems at the LHC energies.

	\begin{acknowledgments}
		We would like to thank Zebo Tang, Wangmei Zha and Qiye Shou for helpful discussions. This work is supported by the National Natural Science Foundation of China (11905188, 11875143 and 12061141008), the National Key Research and Development Program of China (2016YFE0100900) and the Innovation Fund of Key Laboratory of Quark and Lepton Physics LPL2020P01 (LZ).
		
	\end{acknowledgments}


	\bibliography{main}

	\clearpage
	
	\onecolumngrid
	
	\section*{Appendix}
	\begin{table*}[htbp]
		\caption{\label{tabTBW} Charge particle density, extracted kinetic freeze-out parameters and $\chi^{2}/nDoF$ from TBW fits to identified particle transverse spectra in pp collisions at $\sqrt{s}=7$ TeV and 13 TeV, pPb collisions at $\sqrt{s_{NN}}=5.02$ TeV, XeXe collisions at $\sqrt{s_{NN}}=5.44$ TeV, PbPb collisions at $\sqrt{s_{NN}}=2.76$ TeV and $5.02$ TeV with different centralities.}
		\begin{ruledtabular}
			\resizebox{.97\columnwidth}{!}{
				\begin{tabular}{ccccccc}
					
					$\rm system$  & $\rm\ dN/d \eta\;$  & centrality  & $\langle\beta_S\rangle \;$ &  $T \;(\rm{MeV})$  & $ q-1$ & $\chi^{2}/nDoF$ \\
					\hline
					$\rm Pb+Pb $ & $1943$     & $0-5\%$   & $0.908\pm 0.004$ & $94\pm 2$ & $0.018\pm 0.005$ & $324/89$ \\
					$\rm5.02~TeV$   &$1587$     & $5-10\%$  & $0.903\pm 0.003$ & $92\pm 2$ & $0.028\pm 0.004$ & $313/89$ \\
					&$1180$     & $10-20\%$  & $0.895\pm 0.003$ & $95\pm 2$ & $0.029\pm 0.004$ & $323/89$ \\
					&$786$      & $20-30\%$  & $0.874\pm 0.004$ & $96\pm 2$ & $0.039\pm 0.004$ & $267/89$ \\
					&$512$      & $30-40\%$  & $0.838\pm 0.005$ & $95\pm 2$ & $0.055\pm 0.004$ & $212/89$ \\
					&$318$      & $40-50\%$  & $0.779\pm 0.006$ & $93\pm 2$ & $0.074\pm 0.003$ & $182/89$ \\
					&$183$      & $50-60\%$  & $0.692\pm 0.007$ & $93\pm 2$ & $0.090\pm 0.003$ & $183/89$ \\
					&$96.3$     & $60-70\%$  & $0.578\pm 0.014$ & $90\pm 2$ & $0.107\pm 0.003$ & $170/89$ \\
					&$44.9$     & $70-80\%$  & $0.399\pm 0.026$ & $90\pm 2$ & $0.122\pm 0.003$ & $182/89$ \\
					&$17.5$     & $80-90\%$  & $0\pm 0$ & $89\pm 2$ & $0.131\pm 0.002$ & $169/89$ \\
					\hline
					
					$\rm Pb+Pb $ &  $1601$        & $0-0\%$        & $0.887\pm 0.006$      & $96\pm 3$    & $0.016\pm 0.007$ & $157/104$ \\
					
					$\rm2.76~TeV$& $1294$         & $5-10\%$       & $0.877\pm 0.006$    & $96\pm 3$     & $0.024\pm 0.007$ & $151/104$ \\
					
					& $966$          & $10-20\%$     & $0.866\pm 0.007$      & $96\pm 3$    & $0.031\pm 0.007$ & $13/104$ \\
					
					& $537.5$       & $20-40\%$     & $0.820\pm 0.008$      & $98\pm 3$    & $0.049\pm 0.005$ & $133/104$ \\
					
					& $205$          & $40-60\%$     & $0.714\pm 0.012$     & $92\pm 3$   & $0.080\pm 0.004$ & $90/104$ \\
					
					& $55.5$        & $60-80\%$     & $0.488\pm 0.024$      & $94\pm 2$    & $0.105\pm 0.003$ & $134/104$ \\
					\hline
					
					$\rm Xe+Xe $ & $1167$ & $0-5\%$  & $0.899\pm 0.009$ & $89\pm 4$ & $0.032\pm 0.012$ & $140/88$ \\
					
					$\rm5.44~TeV$ & $939$ & $5-10\%$  & $0.889\pm 0.003$ & $88\pm 2$ & $0.041\pm 0.002$ & $135/88$ \\
					
					& $706$ & $10-20\%$  & $0.880\pm 0.010$ & $87\pm 3$ & $0.047\pm 0.010$ & $116/88$ \\
					
					& $478$  & $20-30\%$  & $0.855\pm 0.011$ & $86\pm 3$ & $0.060\pm 0.008$ & $100/88$ \\
					
					& $315$  & $30-40\%$  & $0.809\pm 0.011$ & $83\pm 3$ & $0.079\pm 0.006$ & $85/88$ \\
					
					& $198$  & $40-50\%$  & $0.765\pm 0.013$ & $83\pm 3$ & $0.089\pm 0.005$ & $89/88$ \\
					
					& $118$  & $50-60\%$  & $0.682\pm 0.016$ & $80\pm 2$ & $0.107\pm 0.004$ & $76/88$ \\
					
					& $64.7$  & $60-70\%$  & $0.586\pm 0.022$ & $79\pm 2$ & $0.118\pm 0.004$ & $77/88$ \\
					
					& $22.5$ & $70-90\%$  & $0.412\pm 0.037$ & $80\pm 2$ & $0.127\pm 0.004$ & $86/88$ \\
					\hline

					$\rm p+Pb $ & $45$     & $0-5\%$ & $0.663\pm 0.0130$ & $80\pm 2$ & $0.124 \pm 0.003$ & $232/99$ \\
					
					$\rm5.02~TeV$ & $36.2$ & $5-10\%$  & $0.620\pm 0.015$ & $81\pm 2$ & $0.128\pm 0.003$ & $250/99$ \\
					
					& $30.5$ & $10-20\%$  & $0.574\pm 0.017$ & $82\pm 2$ & $0.130\pm 0.003$ & $254/99$ \\
					
					& $23.2$  & $20-30\%$  & $0.492\pm 0.021$ & $83\pm 2$ & $0.134\pm 0.003$ & $292/99$ \\
					
					& $16.1$  & $30-40\%$  & $0.332\pm 0.035$ & $86\pm 2$ & $0.137\pm 0.003$ & $314/99$ \\
					
					& $9.8$  & $40-60\%$  & $0\pm 0$        & $87\pm 2$ & $0.136\pm 0.001$ &  $345/99$ \\ 
					& $4.4$  & $60-80\%$  & $0\pm 0$        & $90\pm 2$ & $0.119\pm 0.001$ & $447/99$ \\ 
					\hline
					
					$\rm p+p$  & $26.02$  & $0-0.92\%$ & $0.561\pm 0.019$ & $70\pm 2$     & $0.154\pm 0.003$ & $226/89$ \\
					$\rm13~TeV$ & $20.02$ & $0.92-4.6\%$  & $0.493\pm 0.023$ & $73\pm 2$     & $0.151\pm 0.003$ & $271/89$ \\
					& $16.17$ & $4.6-9.2\%$  & $0.414\pm 0.029$    & $76\pm 2$     & $0.149\pm 0.003$ & $288/89$ \\
					& $13.77$ & $9.2-13.8\%$  & $0.329\pm 0.039$   & $79\pm 2$     & $0.148\pm 0.003$ & $308/89$ \\
					& $12.04$ & $13.8-18.4\%$  & $0.236\pm 0.056$  & $80\pm 2$     & $0.148\pm 0.002$ & $316/89$ \\
					& $10.02$ & $18.4-27.6\%$  & $0\pm 0$          & $82\pm 2$     & $0.146\pm 0.002$ & $342/89$ \\
					& $7.95$ & $27.6-36.8\%$  & $0\pm 0$           & $84\pm 2$     & $0.140\pm 0.002$ & $375/89$ \\
					& $6.32$ & $36.8-46.0\%$  & $0\pm 0$           & $85\pm 2$     & $0.134\pm 0.002$ & $436/89$ \\
					& $4.50$ & $46.0-64.5\%$  & $0\pm 0$           & $87\pm 2$     & $0.125\pm 0.001$ & $544/89$ \\
					& $2.55$ & $64.5-100\%$  & $0\pm 0$            & $90\pm 1$     & $0.106\pm 0.001$ & $819/89$ \\  
					\hline
					
					$\rm p+p $ &  $21.3$   & $0-0.95\%$  & $0.513\pm 0.021$ & $75\pm 2$ & $0.144\pm 0.003$ & $257/89$ \\
					$\rm7~TeV$  & $16.5$    & $0.95-4.7\%$ & $0.414\pm 0.026$ & $78\pm 2$ & $0.144\pm 0.003$ & $296/89$ \\
					& $13.5$  & $4.7-9.5\%$ & $0.300\pm 0.039$  & $80\pm 2$ & $0.143\pm 0.002$ & $308/89$ \\
					& $11.5$  & $9.5-14\%$  & $0.155\pm 0.079$  & $82\pm 2$ & $0.143\pm 0.002$ & $325/89$ \\
					& $10.1$  & $14-19\%$  & $0\pm 0$           & $84\pm 2$ & $0.141\pm 0.002$ & $326/89$ \\
					& $8.45$  & $19-28\%$  & $0\pm 0$           & $85\pm 2$ & $0.136\pm 0.002$ & $360/89$ \\
					& $6.72$  & $28-38\%$  & $0\pm 0$           & $89\pm 2$ & $0.123\pm 0.001$ &  $400/89$ \\
					& $5.4$   & $38-48\%$  & $0\pm 0$           & $89\pm 2$ & $0.123\pm 0.001$ & $480/89$ \\
					& $3.9$   & $48-68\%$  & $0\pm 0$           & $92\pm 2$ & $0.114\pm 0.002$ & $523/89$ \\
					& $2.26$  & $68-100\%$ & $0\pm 0$          & $94\pm 2$ & $0.098\pm 0.002$ & $606/89$\\ 
					
				\end{tabular}
			}
		\end{ruledtabular}
	\end{table*}
	
	\begin{table*}[htbp]
		\caption{\label{tabTBW4} Charge particle density, extracted kinetic freeze-out parameters and $\chi^{2}/nDoF$ from TBW4 fits to identified particle transverse spectra in pp collisions at $\sqrt{s}=7$ TeV and 13 TeV, pPb collisions at $\sqrt{s_{NN}}=5.02$ TeV, XeXe collisions at $\sqrt{s_{NN}}=5.44$ TeV, PbPb collisions at $\sqrt{s_{NN}}=2.76$ TeV and $5.02$ TeV with different centralities.}
		\begin{ruledtabular}
			\resizebox{.97\columnwidth}{!}{
				\begin{tabular}{cccccccc}
					
					$\rm system$  & $\rm\ dN/d \eta\;$  & centrality  & $\langle\beta_S\rangle \;$ &  $T \;(\rm{MeV})$  & $ q_M-1$ &$ q_B-1$& $\chi^{2}/nDoF$ \\
					\hline
					$\rm Pb+Pb $ & $1943$     & $0-5\%$   & $0.894\pm 0.005$ & $100\pm 2$ & $0.018\pm 0.006$ & $0.039\pm 0.007$ & $279/88$ \\
					$\rm5.02~TeV$   &$1587$     & $5-10\%$  & $0.894\pm 0.005$ & $97\pm 2$ & $0.026\pm 0.005$ & $0.039\pm 0.005$ & $294/88$ \\
					&$1180$     & $10-20\%$  & $0.887\pm 0.005$ & $98\pm 2$ & $0.028\pm 0.005$ & $0.038\pm 0.005$ & $310/88$ \\
					&$786$      & $20-30\%$  & $0.871\pm 0.005$ & $96\pm 2$ & $0.039\pm 0.004$ & $0.042\pm 0.005$ & $266/88$ \\
					&$512$      & $30-40\%$  & $0.849\pm 0.006$ & $92\pm 2$ & $0.055\pm 0.004$ & $0.048\pm 0.005$ & $203/88$ \\
					&$318$      & $40-50\%$  & $0.805\pm 0.006$ & $87\pm 2$ & $0.075\pm 0.003$ & $0.062\pm 0.004$ & $150/88$ \\
					&$183$      & $50-60\%$  & $0.740\pm 0.008$ & $84\pm 2$ & $0.092\pm 0.003$ & $0.077\pm 0.003$ & $122/88$ \\
					&$96.3$     & $60-70\%$  & $0.674\pm 0.013$ & $77\pm 2$ & $0.110\pm 0.003$ & $0.088\pm 0.00$ & $64/88$ \\
					&$44.9$     & $70-80\%$  & $0.575\pm 0.015$ & $74\pm 2$ & $0.123\pm 0.002$ & $0.098\pm 0.003$ & $51/88$ \\
					&$17.5$     & $80-90\%$  & $0.478\pm 0.031$ & $74\pm 3$ & $0.129\pm 0.002$ & $0.103\pm 0.004$ & $49/88$ \\
					\hline
					
					$\rm Pb+Pb $ &  $1601$        & $0-5\%$       & $0.883\pm 0.009$ & $98\pm 3$ & $0.017\pm 0.009$ & $0.020\pm 0.011$ & $156/103$ \\
					
					$\rm2.76~TeV$& $1294$         & $5-10\%$     & $0.878\pm 0.006$ & $96\pm 3$ & $0.024\pm 0.006$ & $0.023\pm 0.007$ & $151/103$ \\
					
					& $966$          & $10-20\%$    & $0.871\pm 0.009$ & $94\pm 3$ & $0.031\pm 0.007$ & $0.026\pm 0.009$ & $134/103$ \\
					
					& $537.5$       & $20-40\%$    & $0.845\pm 0.008$ & $91\pm 3$ & $0.047\pm 0.005$ & $0.034\pm 0.006$ & $113/103$ \\
					
					& $205$          & $40-60\%$   & $0.770\pm 0.0135$ & $80\pm 3$ & $0.083\pm 0.004$ & $0.063\pm 0.005$ & $55/103$ \\
					
					& $55.5$        & $60-80\%$    & $0.644\pm 0.0177$ & $75\pm 3$ & $0.109\pm 0.003$ & $0.083\pm 0.004$ & $44/103$ \\ 
					\hline
					
					$\rm Xe+Xe $ & $1167$ & $0-5\%$ & $0.892\pm 0.013$ & $90\pm 4$ & $0.033\pm 0.013$ & $0.042\pm 0.017$ & $138/87$ \\
					
					$\rm5.44~TeV$ & $939$ & $5-10\%$  & $0.891\pm 0.012$ & $90\pm 4$ & $0.037\pm 0.011$ & $0.039\pm 0.015$ & $134/87$ \\
					
					& $706$ & $10-20\%$  & $0.882\pm 0.009$ & $87\pm 3$ & $0.046\pm 0.007$ & $0.044\pm 0.009$ & $116/87$ \\
					
					& $478$  & $20-30\%$  & $0.864\pm 0.010$ & $84\pm 3$ & $0.058\pm 0.007$ & $0.050\pm 0.009$ & $97/87$ \\
					
					& $315$  & $30-40\%$  & $0.829\pm 0.014$ & $79\pm 3$ & $0.078\pm 0.007$ & $0.064\pm 0.009$ & $75/87$ \\
					
					& $198$  & $40-50\%$  & $0.791\pm 0.014$ & $78\pm 3$ & $0.088\pm 0.005$ & $0.074\pm 0.007$ & $76/87$ \\
					
					& $118$  & $50-60\%$  & $0.717\pm 0.016$ & $75\pm 3$ & $0.107\pm 0.004$ & $0.094\pm 0.005$ & $63/87$ \\
					
					& $64.7$  & $60-70\%$ & $0.656\pm 0.022$ & $70\pm 3$ & $0.120\pm 0.004$ & $0.101\pm 0.005$ & $47/87$ \\
					
					& $22.5$ & $70-90\%$  & $0.547\pm 0.030$ & $69\pm 3$ & $0.129\pm 0.004$ & $0.106\pm 0.004$ & $37/87$ \\
					\hline

					$\rm p+Pb $ & $45$     & $0-5\%$& $0.768\pm 0.013$ & $65\pm 3$ & $0.126\pm 0.004$ & $0.091\pm 0.005$ & $88/98$ \\
					
					$\rm5.02~TeV$ & $36.2$ & $5-10\%$  & $0.740\pm 0.012$ & $65\pm2 $ & $0.130\pm 0.003$ & $0.096\pm 0.005$ & $93/98$ \\
					
					& $30.5$ & $10-20\%$ & $0.710\pm 0.013$ & $66\pm 2 $ & $0.132\pm 0.003$ & $0.100\pm 0.004$ & $91/98$ \\
					
					& $23.2$  & $20-30\%$  & $0.673\pm 0.016$ & $64\pm 3$ & $0.137\pm 0.003$ & $0.103\pm 0.004$ & $76/98$ \\
					
					& $16.1$  & $30-40\%$  & $0.604\pm 0.016$ & $65\pm 2$ & $0.138\pm 0.002$ & $0.105\pm 0.003$ & $69/98$ \\
					
					& $9.8$  & $40-60\%$  & $0.507\pm 0.027$ & $65\pm 3$ & $0.140\pm 0.002$ & $0.107\pm 0.003$ & $37/98$ \\
					& $4.4$  & $60-80\%$ & $0.279\pm 0.046$ & $71\pm 2$ & $0.134\pm 0.002$ & $0.104\pm 0.003$ & $21/98$ \\ 
					\hline
					
					$\rm p+p$  & $26.02$  & $0-0.92\%$ & $0.678\pm 0.016$ & $59\pm 3$ & $0.154\pm 0.003$ & $0.119\pm 0.005$ & $102/88$ \\
					$\rm13~TeV$ & $20.02$ & $0.92-4.6\%$  & $0.647\pm 0.015$ & $60\pm 2$ & $0.152\pm 0.003$ & $0.113\pm 0.004$ & $89/88$ \\
					& $16.17$ & $4.6-9.2\%$   & $0.608\pm 0.019$ & $61\pm 3$ & $0.150\pm 0.003$ & $0.111\pm 0.004$ & $80/88$ \\
					& $13.77$ & $9.2-13.8\%$  & $0.576\pm 0.021$ & $63\pm 3$ & $0.148\pm 0.003$ & $0.109\pm 0.004$ & $72/88$ \\
					& $12.04$ & $13.8-18.4\%$  & $0.548\pm 0.017$ & $63\pm 2$ & $0.147\pm 0.002$ & $0.109\pm 0.003$ & $64/88$ \\
					& $10.02$ & $18.4-27.6\%$ & $0.511\pm 0.026$ & $64\pm 3$ & $0.145\pm 0.002$ & $0.107\pm 0.004$ & $56/88$ \\
					& $7.95$ & $27.6-36.8\%$  & $0.453\pm 0.024$ & $66\pm 2$ & $0.143\pm 0.002$ & $0.105\pm 0.003$ & $43/88$ \\
					& $6.32$ & $36.8-46.0\%$  & $0.399\pm 0.029$ & $68\pm 2$ & $0.140\pm 0.002$ & $0.103\pm 0.003$ & $35/88$ \\
					& $4.50$ & $46.0-64.5\%$  & $0.284\pm 0.044$ & $71\pm 2$ & $0.136\pm 0.002$ & $0.100\pm 0.003$ & $323/88$ \\
					& $2.55$ & $64.5-100\%$  & $0\pm 0$ & $76\pm 1$ & $0.121\pm 0.001$ & $0.088\pm 0.001$ & $80/88$ \\  
					\hline
					
					$\rm p+p $ &  $21.3$   & $0-0.95\%$  & $0.678\pm 0.012$ & $60\pm 1$ & $0.144\pm 0.002$ & $0.105\pm 0.004$ & $100/88$ \\
					$\rm7~TeV$  & $16.5$    & $0.95-4.7\%$ & $0.631\pm 0.018$ & $61\pm 2$ & $0.143\pm 0.002$ & $0.105\pm 0.004$ & $98/88$ \\
					& $13.5$  & $4.7-9.5\%$ & $0.591\pm 0.016$ & $62\pm2$ & $0.142\pm 0.002$ & $0.104\pm 0.003$ & $84/88$ \\
					& $11.5$  & $9.5-14\%$  & $0.560\pm 0.018$ & $63\pm 2$ & $0.140\pm 0.002$ & $0.102\pm 0.003$ & $74/88$ \\
					& $10.1$  & $14-19\%$  & $0.532\pm 0.025$ & $64\pm 3$ & $0.139\pm 0.002$ & $0.102\pm 0.003$ & $65/88$ \\
					& $8.45$  & $19-28\%$  & $0.498\pm 0.028$ & $65\pm 3$ & $0.137\pm 0.002$ & $0.100\pm 0.003$ & $52/88$ \\
					& $6.72$  & $28-38\%$  & $0.444\pm 0.022$ & $671\pm 2$ & $0.135\pm 0.001$ & $0.099\pm 0.002$ & $36/88$ \\
					& $5.4$   & $38-48\%$  & $0.385\pm 0.047$ & $69\pm3$ & $0.132\pm 0.002$ & $0.097\pm 0.003$ & $26/88$ \\
					& $3.9$   & $48-68\%$  & $0.253\pm 0.120$ & $73\pm 5$ & $0.129\pm 0.002$ & $0.097\pm 0.005$ & $18/88$ \\
					& $2.26$  & $68-100\%$ & $0     \pm 0$ & $75\pm 2$ & $0.117\pm 0.002$ & $0.086\pm 0.001$ & $42/88$ \\

				\end{tabular}
			}
		\end{ruledtabular}
	\end{table*}

\end{document}